# Managing Uncertainty in Teacher Vibe Coding: A Comparison of Teacher and Expert Workflows

Jiaju Lin and Ellen Wenting Zou

Pennsylvania State University

## Abstract

Vibe coding, or natural language software development, enables teachers to become creators of digital teaching artifacts. However, novice educators oftentimes face different barriers. This qualitative study examined the workflows of pre-service teachers, in-service teachers, and expert developers through think-aloud protocols. Findings reveal vibe coding operates through a loop of problem framing, development, and verification governed by uncertainty management. Teachers reduced uncertainty by grounding designs in authentic needs: pre-service teachers drew on pedagogical principles, whereas in-service teachers relied on everyday teaching challenges. Conversely, Experts reduced ambiguity by constraining requirements, decomposing artifacts into testable modules, and automating tests. These differences reveal complementary forms of pedagogical, contextual, and technical expertise. Future work may be about supporting nontechnical teachers in developing reliable and educational applications.

## Introduction

Teachers are increasingly understood not only as implementers of instructional materials but also as designers of learning experiences and technology-rich activities (Warr & Mishra, 2021). Research on teachers as designers indicates that active participation in educational technology design can support the development of technological pedagogical knowledge, strengthen ownership, and improve the practicality of technology-enhanced activities in classroom contexts (Koehler & Mishra, 2005; Cviko et al., 2014).

Large language model–based code generation (or "vibe coding") may extend this design role by allowing teachers to create functional applications through natural-language interaction rather than conventional programming. However, removing the need to write code syntax does not eliminate the expertise required to develop reliable software. Research with novice and end-user programmers has identified an “abstraction gap” between users’ intended outcomes and the forms of requests that code-generating models can interpret and implement effectively (Liu et al., 2023). Novices also experience difficulties evaluating generated code, diagnosing errors, regulating their reliance on AI, and modifying initially successful solutions when requirements change (Prather et al., 2023; Kazemitabaar et al., 2023).

Existing research establishes important challenges in AI-assisted programming but lacking in explaining how teachers use emerging “vibe-coding” tools to design educational applications. Addressing this gap requires comparative evidence that distinguishes barriers associated with limited software-development expertise from design practices grounded in teachers’ pedagogical expertise, thereby clarifying how educational application development differs from general-purpose AI-assisted programming.

This qualitative comparative study therefore examines the vibe-coding workflows of pre-service teachers, in-service teachers, and experienced AI-assisted developers. By identifying where these workflows converge and diverge, the study examines how pedagogical, contextual, and software-development expertise shape vibe-coding practice. Specifically, it addresses the following research question:

> How do vibe-coding workflows converge and diverge across pre-service teachers, in-service teachers, and experienced AI-assisted developers?

The study contributes an empirically grounded account of teacher vibe coding and identifies practices that may inform future instructional support for nontechnical educators.

# Conceptual Perspective

This study adopts a teacher-as-designer perspective, which conceptualizes teachers as active designers of learning experiences and technology-rich activities rather than merely as implementers of instructional technologies (Warr & Mishra, 2021; Cviko et al., 2014). Teacher design is situated in pedagogical intentions, classroom routines, learner needs, and anticipated conditions of implementation. Drawing additionally on Schön’s (1983) account of reflective practice, design is understood as an iterative conversation with an evolving situation:

practitioners frame a problem, make provisional moves, interpret the consequences, and revise either the artifact or their understanding of the problem. This perspective directs attention to how participants frame educational needs, respond to AI-generated prototypes, and evaluate both technical performance and contextual appropriateness. The Replacement–Amplification –Transformation framework (Hughes et al., 2006) is used separately as an artifact-level lens to characterize the educational functions represented in teacher-designed applications.

# Methodology

This qualitative comparative study examined vibe-coding workflows among 15 pre-service teachers, 13 in-service teachers, and 10 expert developers. Both teacher groups had no programming background, whereas experts regularly used AI coding assistants and served as a benchmark for mature uncertainty-management practices.

Participants completed remote Zoom sessions including background, interviews, a 30-minute development task, and post-task debriefing. Teachers developed self-defined applications, while experts implemented clear-defined requirements generated by teachers. This design ensured that the experts worked on authentic educational problems and facilitated comparison across participant groups. Teachers used a common AI coding platform, whereas experts used familiar tools. All participants shared their screens and thought aloud.

Data included questionnaires, interviews, screen recordings, verbalizations, and coding traces. We used inductive thematic analysis to discover the patterns in participants workflows.

The teacher-built artifacts were coded using the Replacement-Amplification-Transformation (RAT) framework to characterize the integration of technology into the designed learning activities. Expert-produced artifacts were not coded separately because they were developed from teacher-generated requirements and were treated as matched counterparts to the corresponding teacher-produced artifacts. Two trained annotators independently coded each artifact. Inter-rater agreement exceeded 90%, and disagreements were resolved through discussion and consensus.

# Findings

We observed pre-service teachers (PST01–PST15), in-service teachers (IST01–IST13), and coding experts (EXP01–EXP10) as vibe–coded educational web apps. The work tended to follow a loop of Problem framing, Development, Verification, under the governance of uncertainty management. Across participant groups, we observed different focus on the loop stages and uncertainty management methods.

## 5.1 Problem framing stage: from teacher struggles to AI-negotiated requirements

Participants initiated their artifact design by framing an ambiguous problem into a concrete requirement. Our observation showed that it involved reflecting daily teaching struggles and constraints, then turning them into artifact design. Participants modified their specific requirements to a clearer and concrete design representation with decreased uncertainty. Each group differed in tasks they wanted to solve, contexts they cared about, and their perceptions of the capability and constraints of vibe-coding.

Teachers started from a teaching struggle, while participant groups' understanding of struggle varies. RAT analysis further revealed different uses of technology. In-service teachers primarily produced replacement artifacts that automated familiar routines, such as weekly reading checks with automated assessment, tablet-based worksheets, and tools fitting “the lesson I already teach” (IST03; Figure 2). Pre-service teachers more often pursued amplification or transformation, drawing on pedagogical principles to design mission-based games, immediate reading assistant (PST04; Figure 3), and a collaborative historical modeling game (PST10; Figure 4).

Teachers' initial ideas of addressing teaching struggles were not always directly translated into the requirement towards AI. It may be shaped by beliefs about their vibe-coding abilities. PST01 envisioned an AI tutor for every student but produced a complex multi-module course, while IST03's textbook-translation idea became a multilingual course-selection page without textbook integration. In post-experience interview, they admitted that ‘I think that (initial design) may be too complex and out of the scope of vibe-coding ', ‘Since I have no idea how to build an AI tutor and test that.' In other words, participants discarded the design they were uncertain about.

As part of the vibe-coding loop, problem framing did not always precede the first round of prompting in a linear sequence. Instead, problem framing and artifact design were refined through prototype-mediated reflection towards higher certainty: prototypes materialized tentative solutions while eliciting more tacit knowledge about the relevant contexts and constraints. Many teachers reported that their requirements became concrete only after they saw a functioning webpage. Some (PST07, IST04) even said AI's suggestions were beyond expectation.

There were also differences in elicited tactic contexts and constraints. In-service teachers expanded problems toward classroom workflows: IST11 extended a student quiz to support before-, during-, and after-class decisions, including identifying struggling students, error patterns, and next instructional steps. Pre-service teachers focused more on existing feature improvements, such as voice and badge colors.

Experts instead translated teacher-generated ideas into clearer specifications to reduce uncertainty associated with black-box AI, using ambiguity checks, success criteria, and feasibility assessments. This process involved iterative negotiation with the AI. As EXP02 noted, asking the AI to pose clarifying questions also helped experts refine their own understanding of the problem.

## 5.2 Development stage: micro-experimentation and the limits of supervision

In the development stage, participants confronted two related uncertainties: whether the AI was capable of implementing a particular requirement and whether the growing artifact remained aligned with the intended design. To address them, participants mainly used two strategies: testing AI's ability by micro-experiment and supervising the artifact as it grew. Teachers and experts participated in both, but the balance between them differed.

Many participants, including several experts, were uncertain about AI's capabilities in relatively specialized tasks, such as implementing a particular technology stack or applying curriculum standards. To manage this uncertainty, they often engaged in a micro-experimentation loop, testing the AI on a narrowly defined subtask before full-scale implementation. For example, PST06 conducted a focused probe of the AI's domain knowledge: "Let's test whether it knows the Grade 5 mathematics curriculum and can generate the instructional content by itself." Experts used a more rigorous version . As EXP03 explained, "I will ask the AI to complete task one first to see whether it can build Swift apps." This approach enabled experts to identify errors at an early stage and update their understanding of the AI's limitations.

The ability to supervise artifact construction was largely limited to experts. Although some teachers attempted to understand the underlying code by examining its architecture, they were often overwhelmed by its complexity. Experts, similarly, did not read code line by line. They decomposed the artifact into functional modules and tested separately. Some experts (EXP01, EXP03, and EXP06) defined explicit success criteria for each stage of construction during the problem-framing phase and oversaw them when developing. As EXP06 said, "Detailed success criteria make the planning phase complex but make me better understand the project.'

Through micro-experimentation and structured process supervision, participants developed an understanding of what the AI could build and what it had produced, thereby reducing uncertainty during the development stage.

## 5.3 Verification stage: functional checks versus pedagogical judgment

During verification, participants evaluated the quality of the completed artifact to mitigate potential risks in real-world use. By checking functionality, applying professional judgment, and using prototype-driven reflection, participants reduced uncertainty about the artifact's future implementation and returned to the problem-framing stage when necessary.

Teachers checked functions lightly and judged situated quality carefully. Teachers' test for their artifacts was almost manually, which limited the coverage. PST04: "I didn't try all the wrong answers. I just checked that the first question can be moved to the next one." By contrast, teachers drew sharp judgment on the pedagogical quality. IST06 rejected red feedback as too error-like for children, and PST12 requested slower animations and more encouraging responses.. In-service teachers were more sensitive about classroom feasibility. IST07 redefined the artifact from an in-classroom tool to an after-school work, driven by the concern on the instability of technologies. “ My children are young, they will lose patience if the latency is high“. Some participants even preferred “boring but reliable” traditional tools because established classroom routines with minimum uncertainty in implementation.

Experts used more systematic verification, like AI-generated test cases covering common interaction paths and code-quality review. These tests reduced implementation uncertainty by covering everyday user interaction paths. However, polished AI-generated code made hidden errors harder to detect. EXP08 complained that “Previously, I can allocate my attention to review poor-quality code. But now, all codes are so clean and formal... “. Experts therefore focused on defining appropriate quality gates and revised these criteria when failures emerged, linking verification back to problem framing, which was the key method used to improve certainty in code quality.

## Conclusion

The growing accessibility of AI code generation creates new opportunities for teachers to design customized educational applications without conventional programming expertise. However, the expertise required for successful educational vibe-coding remains poorly understood in existing literature. We address this gap through a qualitative comparative study of three participant groups, examining how technical and pedagogical expertise shape vibe-coding workflows. From these comparisons, we derive an initial conceptualization of the vibe-coding paradigm for educational application development and identify implications for preparing teachers to become AI-assisted educational designers.

Across groups, vibe coding followed an uncertainty-managed loop of problem framing, development, and verification. Pre-service teachers contributed pedagogical improvement and creativity, in-service teachers focused on contextual classroom knowledge, and developers emphasized technical strategies for requirement clarification, modularization, and verification. Effective educational vibe coding therefore integrates pedagogical, contextual, and technical expertise.

Future research should translate these workflow findings into instructional sequences and scaffolding strategies and evaluate whether such supports improve both the technical reliability and pedagogical value of teacher-built applications.

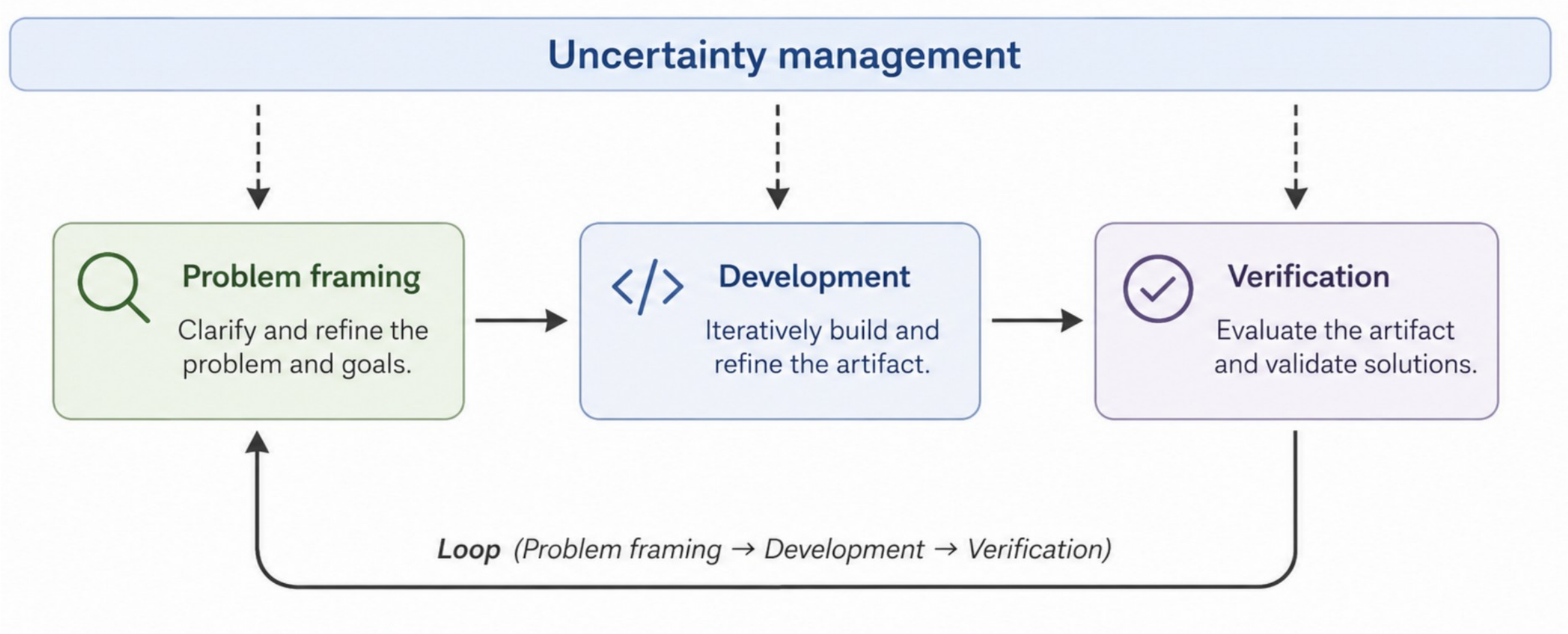

Figure 1. The vibe-coding loop with uncertainty management.

| Group | Replacement (R) | Amplification (A) | Transformation (T) |
|---|---|---|---|
| Pre-service teachers (*n* = 15) | 5/15 | 9/15 | 1/15 |
| In-service teachers (*n* = 13) | 8/13 | 5/13 | 0/13 |

Table 1 Artifact distribution under the RAT framework. Expert artifacts were treated as matched counterparts because they were developed from teacher-generated requirements.

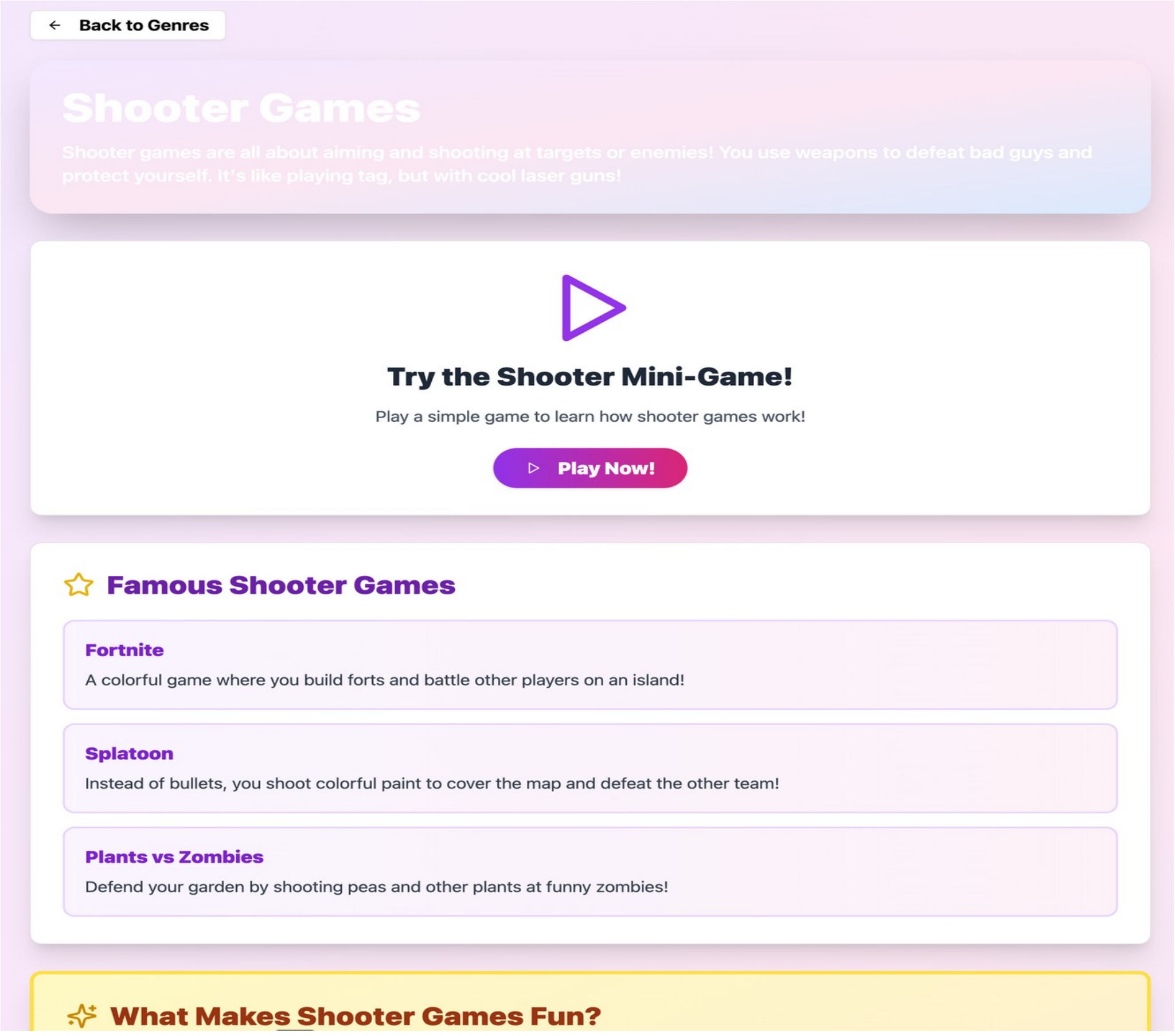

Figure 2. The screenshot of a game design gallery from IST03 (replacement level) .

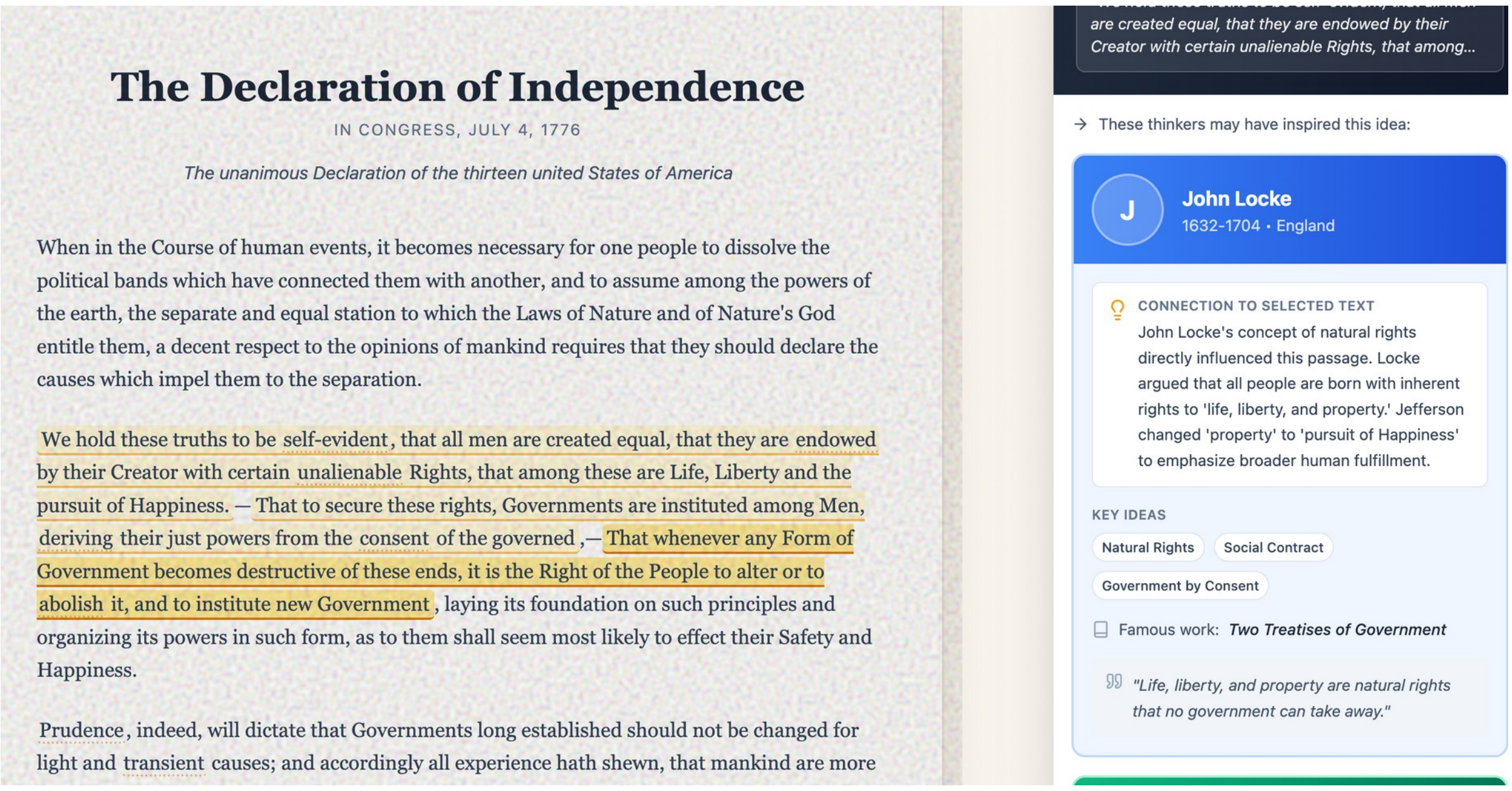

Figure 3. The screenshot of a reading assistant from PST04 (augmentation level) .

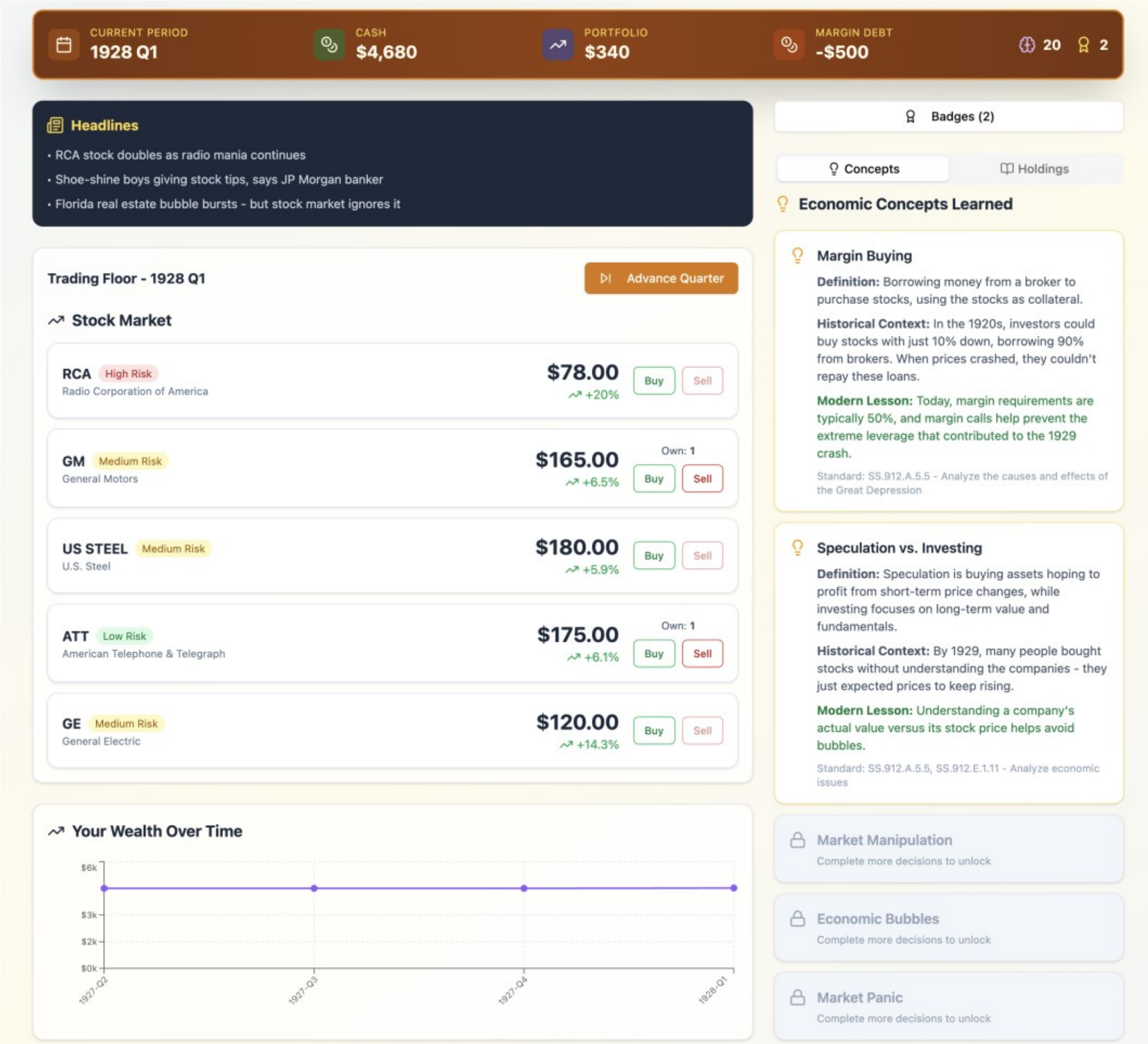

Figure 4. The screenshot of a collaborative game from PST10 to simulate the great crash (transformation level) .